# Energy Efficient MAC Protocols in Wireless Body Area Sensor Networks - A Survey

N. Javaid[1], S. Hayat[1], M. Shakir[1], M. A. Khan[1], S. H. Bouk[1], Z. A. Khan[2]

[1]*COMSATS Institute of Information Technology, Islamabad, Pakistan.*

*[nadeemjavaid@comsats.edu.pk]*

[2]*Faculty of Engineering, Dalhousie University, Halifax, Canada.*

*Abstract*

**In this paper, we first presented an analytically discussion about energy efficiency of Medium Access Control (MAC) protocols for Wireless Body Area Sensor Networks (WBASNs). For this purpose, different energy efficient MAC protocols with their respective energy optimization techniques; Low Power Listening (LPL), Scheduled Contention and Time Division Multiple Access (TDMA), are elaborated. We also analytically compared path loss models for In-body, On-body and Off-body communications in WBASNs. These three path loss scenarios are simulated in MATLAB and results shown that path loss is more in In-body communication because of less energy level to take care of tissues and organs located inside human body. Secondly, power model for WBASNs of Carrier Sense Multiple Access with Collision Avoidance (CSMA/CA) and beacon mode is also presented. MATLAB simulations results shown that power of CSMA/CA mode is less as compared to beacon mode. Finally, we suggested that hybrid mode is more useful to achieve optimization in power consumption, which consequently results in high energy efficiency.**

## I. INTRODUCTION

Evolution of wireless, medical and computer networking technologies have merged into an emerging horizon of science and technology called Wireless Body Area Networks (WBANs). However, applications of WBASNs are not limited to medical field only. WBASN is also considered as an important branch of WSN due to its appliances. In both WBASN and WSN, energy efficiency, mobility and localization of sensor snodes is an eye-catching issue to achieve better optimization of WSNs. Literature [1] proposes a learning based algorithm to achieve localization of sensor nodes. In WBANs, Miniaturization and connectivity are notable parameters. WBANs consist of three levels; first level is low power sensors or nodes which are battery powered and need to be operated for a long time without recharging and battery replacement. These nodes may be placed on the body, around the body or implanted in the body. Second level is called Master Node (MN), gateway or coordinator which controls its child nodes; its power requirements may be less strengthened than nodes due to its applications and flexibility. Third level is the local or metropolitan or internet network that serves for monitoring purposes. Energy efficiency or effective power consumption of a system is one of the basic requirements for WBANs because of limited power of batteries. In [2], [3] several issues are considered for increasing the energy efficiency of the network, but the most suitable layer for discussing energy and power issues is MAC layer. The basic way of saving power or enhancing energy efficiency is to minimize the energy wastage. There are several sources of energy wastage including packet collision, over hearing, idle listening, control packet overhead, etc. Major source of energy inefficiency among the above listed sources is packet collision for WBANs. Collision avoidance for energy efficiency, minimum latency, high throughput, and communication reliability, are basic requirements in the design of MAC protocol.

The fundamental way of saving power or enhancing energy efficiency is to minimize the energy wastage. Simulations are performed in MATLAB for different scenarios to compute path loss. Results show that path loss is maximum in In-body communication, as compare to On-body and Off-body communication because human body is composed of tissues and organs which are sensitive for alectromegnatic radiations of transciever leads to complexities. Therefore, communication becomes difficult and results in high path loss. On-body and Off-body also result some variations when the source and destination sensors or nodes are placed Line of Sight (LoS) and Non Line of Sight (NLoS). In this paper, we therefore, provide a survey of energy efficient MAC protocols for WBANs. Sources that contribute to the energy inefficiency in a particular protocol is also identified. Path loss for WBANs is described in detail. Finally, power model of beacon and Multiple Access with Collision Avoidance (CSMA/CA) mode for WBANs is discussed.

## II. RELATED WORK

Gopalan et al. [4] survey MAC protocols for WBANs along with the comparison of four protocols i.e., Energy Efficient MAC, MedMac, Low Duty Cycle MAC, and Body MAC. Some key requirements and sources of energy wastage are also discussed. They also discussed some open research issues in this survey. Still a lot of work has to be done in data link layer, network layer and cross layer design. In [5], Barati has deigned an efficient real time and error control algorithm to enhance network lifetime of wireless sensor networks using redundant residue number system. In [6], Kutty, S et al. discusses the design challenges for MAC protocols for WBANs. They classify data traffic for WBANs into three categories: energy minimization techniques, frame structures and network architecture. However the comparison of protocols is not provided by them. Sana Ullah et al. in [7] provide relatively a comprehensive study of MAC protocols for WBANs. Comparison of the low power listening, scheduled contention and Time Division Multiple Access (TDMA) is provided. MAC requirements, frame structures and comparison of different protocols and their trade-offs are discussed in detail.

## III. ENERGY MINIMIZATION TECHNIQUES IN MAC PROTOCOLS FOR WBASNs

Low power utilization mechanisms play an important role in performance enhancement of MAC protocol. In this section, different approaches and techniques that provide energy efficiency in MAC protocols for WBASNs are discussed and compared.

Energy efficiency is an important issue because the power of sensor nodes in WBANs is limited and long duration of operation is expected. The key concept for low power consumption is to minimize the energy consumption in the following sources: sensing, data processing and communication.

Most of the energy wastage is caused during communication process because of the collision of packets, idle listening, over hearing, over-emitting, control packet overhead and traffic fluctuations. Idle listening can be reduced through duty cycling. To reduce energy waste in order to increase network's life time and to enhance the performance of MAC protocol, different wake-up mechanisms are used.

There are three main approaches adopted for the energy saving mechanisms in MAC protocols for WBANs; Low Power Listening (LPL), Scheduled Contention, TDMA.

### A. LPL

LPL procedure is that "node awakes for a very short period to check activity of channel". If the channel is not idle then the node remains in active state to receive data and other nodes go back to sleeping mode. This is also termed as channel polling [7]. This procedure is performed regularly without any synchronization among the nodes. A long preamble is used by the sender to check polling of the receiver. LPL is sensitive to traffic rates which results in degradation of performance in the scenario of highly varying traffic rates. However, it can be optimized effectively for already known periodic traffic rates. Wise-MAC [7] is one of the MAC protocols, which is based on LPL. This protocol reduces Idle listening using non-persistent CSMA and preamble sampling technique.

### B. Scheduled Contention

Scheduled Contention is the combination of the scheduling and contention based mechanisms to effectively cope with the scalability and collision problems. In contention based protocols, contending nodes try to access the channel for data transmission therefore, probability of packet collision is greatly increased. Example of contention based MAC protocol is Carrier Sense Multiple Access/Collision Avoidance (CSMA/CA) in which Clear Channel Assessment (CCA) is performed by the nodes before transmitting data. Scheduling or contention free means that each node has the schedule of transmission in the form of bandwidth or time slot assignment. TDMA, Carrier

Sense Multiple Access (CDMA) and Frequency Division Multiple Access (FDMA) schemes are some examples of scheduling mechanisms. However, CDMA and FDMA are not suitable for WBANs because of high computational overhead and frequency limitations, respectively.

Sensor MAC (S-MAC) is one of a MAC protocol based on the scheduled contention. In this protocol, low duty mode is set as default mode for all the nodes which assures the coordinated sleeping among neighboring nodes. The energy wastage due to collision, overhearing, idle listening etc. is minimized because the node is turned on only for transmission of data and remains in sleep mode, otherwise.

### C. TDMA

TDMA is the most suitable scheduling scheme, even though it requires extra power consumption due to its sensitivity for synchronization. The scheduled contention is the combination of scheduling and contention based mechanisms. In scheduled contention, a common schedule is adopted by all the nodes to transmit data. This schedule is exchanged periodically among the nodes to make communication adaptive, flexible and scalable.

In TDMA mechanism, a super frame consists of a fixed number of time slots is used. Time slots are allocated to the sensor nodes by a central node and are known as Master Node (MN), Cluster Head (CH), coordinator or Base Station (BS). Traffic rate is one of the key parameter used by the coordinator to allocate time for each contending node. This scheme is highly sensitive to clock drift, which may result in limited throughput. The scheme is power efficient because a node gets time slot for transmission of data and remains in sleep mode for rest of the time. However, the synchronization requirements may degrade performance in terms of power consumption. Preamble-Based TDMA (PB-TDMA) protocol is one of the TDMA based protocol. Other examples include Body-MAC (B-MAC) [8], Med MAC [7] etc.

## IV. ENERGY EFFICIENT MAC PROTOCOLS

In this section, we briefly discuss the energy efficient MAC protocols for WBAN.

### A. Okundu MAC Protocol

An energy efficient MAC protocol for single hop WBANs is proposed by Okundu *et al.* in [9]. This protocol consists of three main processes: link establishment, wakeup service, and alarm process. Basic energy saving mechanism of this protocol consists of central control of wakeup/sleep time and Wakeup Fall-back Time (WFT) processes. WFT mechanism is used to avoid collision due to continuous time slot. This mechanism states that, if a slave node wants to communicate with a MN and it fails in its task due to MN's other activities, then it goes back to sleep mode for a specific time computed by WFT. However, data is continuously being buffered during the sleep time.

To minimize time slot collision, the concept of WFT has been introduced. This concept helps every slave node to maintain a guaranteed time slot even if it fails to communicate with the MN. In this protocol, problems like idle listening and over-hearing can be reduced because of central management of traffic.

In one cluster, only 8 slave nodes can be connected to MN, which restricts inclusion of other slave nodes. In link establishment, wakeup service, and alarm processes, communication is initiated by the MN. Another main problem is that, only one slave node can join network at a time.

### B. Med Mac Protocol

N. F. Timmons *et al.* in [8] propose a TDMA-Based MAC protocol for WBANs called Med MAC. The protocol consists of two schemes for the power saving: Adaptive Guard Band Algorithm (AGBA) and with Drift Adjustment Factor (DAF). AGBA along with time stamp is used for synchronization among coordinator and other nodes. This synchronization is introduced using Guard Band (GB) between time slots to allow the node to sleep for many beacon periods. GB is used to compensate drift due to clocks. DAF is used to minimize bandwidth. GB is calculated by AGBA and shows the worst cases. However, practically gaps may be different between time slots depending upon application scenarios. DAF adjusts GB according to practical situation and avoids overlapping between consecutive slots. MedMac outperforms IEEE 802.15.4 for Class *0* (lower data rate applications such as health monitoring and fitness) and Class *1* (medium data rate medical applications such as EEG). Energy waste due to collision is reduced by introducing Guaranteed Time Slot (GTS). Each device has exclusive use of a channel for a fixed time slot, therefore, synchronization overhead is also reduced. This protocol works efficiently for low data rate applications, and work inefficiently for high data rate applications. However, In-body and On-body applications of WBAN are usually of higher data rate.

### C. Low Duty Cycle MAC Protocol

Low Duty Cycle MAC protocol for WBANs is designed in [10]. In this protocol, analog to digital conversion is performed by slave nodes while the other complex tasks such as digital signal processing is carried out at MN. MNs are supposed to be less power than slave nodes. This protocol introduces the concept of Guard Time ($T_g$) to avoid overlapping between consecutive time slots. After T frames a Network Control (NC) packet is used for general network information. Power saving is achieved by using effective TDMA strategy.

This protocol is energy efficient because it sends data in short bursts. By using TDMA strategy, this protocol effectively overcomes the collision problem. It allows monitoring patient's condition and can reduce the work load on medical staff, while keeping minimum power usage. As, TDMA strategy is used, and it is found that TDMA is more suitable for static type of networks with a limited number of sensors generating data at a fixed rate therefore, this protocol may not respond well in a dynamic topology.

### D. B-MAC Protocol

B-MAC protocol achieves energy efficiency by using three bandwidth management schemes: Burst, Periodic, and Adjust bandwidth. Burst bandwidth consists of temporary period of the bandwidth, which includes several MAC frames and recycled by the gateway (coordinator). Bandwidth is reduced to half if it does not fully utilized by the nodes, which is also informed about reduction of bandwidth. Periodic bandwidth is a provision for a node to have access to the channel exclusively within a portion of each MAC frame or few MAC frames. It is also allocated by the gateway based on node's QoS requirements and current availability of the bandwidth [11]. Adjust bandwidth defines the amount of bandwidth to be added to or reduced from previous Periodic Bandwidth [11]. Sensor nodes can enter into sleep mode and wake up only when they have to receive and transmit any data to the gateway, because the nodes and gateway are synchronized in time. Time slot allocation in Contention Free Period (CFP) is collision free, which improves packet transmission and thus, saves energy. This protocol uses CSMA/CA in the uplink frame of Contention Access Period (CAP) period, which is not reliable scheme due to its unreliable CCA and collision issues.

### E. Ta-MAC Protocol

Ta-MAC [10] protocol utilizes traffic information to enable low-power communication. It introduces two wakeup mechanisms: a traffic-based wakeup mechanism, and a wakeup radio mechanism. Former mechanism accommodates normal traffic by exploiting traffic patterns of nodes, whereas, later mechanism accommodates emergency and on-demand traffic by using a wakeup radio signal. In the traffic-based wakeup mechanism, the operation of each node is based on traffic patterns. The initial traffic pattern is defined by the coordinator and can be changed later. The traffic patterns of all nodes are organized into a table called traffic-based wakeup table. In wakeup radio mechanism, a separate control channel is used to send a wakeup radio signal. The coordinator and the member node send wakeup radio signal in on-demand and emergency case.

In Ta-MAC, a node wakes up whenever, it has a packet to send/receive. Since the traffic patterns are pre-defined and known to the coordinator, it does not have to wait for resource allocation information/beacon. As a result, delay is minimized comparative to other MAC protocols. This protocol accommodates normal, emergency and on-demand traffic in a reliable manner. To achieve energy efficiency in MAC protocol, the central coordination and resource allocation is based upon the traffic patterns of the nodes.

As, in this protocol, the traffic pattern are defined by the coordinator, in a static topology. Therefore, it does not work efficient in dynamic topology (in dynamic topology, traffic patterns are changed frequently).

### F. S-MAC Protocol

S-MAC [11] is proposed for WBASNs. The protocol uses fixed duty cycles to solve idle listening problem. Nodes wakeup after a specific time, as assigned by coordinator, sends data and goes back to sleep mode again. As, all the nodes are synchronized, therefore, collision can also be easily avoided. S-MAC gives considerably low latency. In this protocol, time synchronization overhead may be prevented due to sleep schedules.

Fluctuating traffics are not supported and no priority is given to the emergency traffic scenarios by S-MAC. Therefore, it is not a reliable for WBANs. Overhearing and collision may occur if the packet is not destined to the listening node.

### G. T-MAC Protocol

Mihai *et al.* [14] suggested Time-out MAC (T-MAC) for WBASNs. It uses flexible duty cycles for increasing energy efficiency. In T-MAC, the node wakes up after time slot assignment, send pending messages. If there is no activation event for Time Interval (TA), the node goes back to sleep mode again. If a node sends Route To Send (RTS) and does not receive Clear To Send (CTS), then sends RTS two more times before going to sleep. To solve early sleep problem, it uses future RTS for taking priority on full buffer.

In T-MAC, packets are sent in burst, as a result delay is minimized. It also outperforms other MAC protocols under variable load. The main disadvantage in this protocol is that it suffers from sleeping problems.

### H. H-MAC Protocol

H-MAC uses Heart Beat Rhythm information for synchronization of nodes. This avoids the use of external clock and thus reducing the power consumption. Also Guaranteed Time Slot (GTS) provision to each node helps to avoid collision. H-MAC aims to improve BSNs energy efficiency by exploiting heartbeat rhythm information, instead of using periodic synchronization beacons to perform time synchronization [7].

Although, H-MAC protocol reduces extra energy cost of synchronization, however, it does not support sporadic events. Since TDMA slots are dedicated and are not traffic adaptive, H-MAC protocol encounters low spectral/bandwidth efficiency in case of low traffic. The heartbeat rhythm information varies depending on patient's condition. It may not reveal valid information for synchronization all the time [7].

### I. DTDMA Protocol

Reservation based dynamic TDMA (DTDMA) protocol uses slotted ALOHA in CAP field of super frame to reduce collisions and to enhance power efficiency.

Through the adaptive allocation of the slots in a DTDMA frame, WBAN's coordinator adjusts the duty cycle adaptively with traffic load. Comparing with IEEE 802.15.4 MAC protocol, DTDMA provides more dependability in terms of lower packet dropping rate and low energy consumption especially for an end device of WBAN [5]. It does not support emergency and on-demand traffic. Furthermore, DTDMA protocol has several limitations when considered for the Medical Implant Communication Service (MICS) band. The MICS band has ten sub-channels and each sub-channel has 300 Kbps bandwidth. DTDMA protocol can operate on one sub-channel, however, cannot operate on ten sub-channels simultaneously [7]. The main purpose of a MAC protocol is to provide energy efficiency, network stability, and bandwidth utilization and reduce packet collision. Energy efficiency of MAC protocols with respect to path loss is discussed in this section.

## V. PATH LOSS MODEL FOR WBASNs

Path loss in WBASNs is also of great importance. It is due to reflection, diffraction, scattering and shadowing of signals from, tissues and organs located inside the human body and objects in the surroundings. Large scale fading of signal is also known as path loss. Reduction in path loss will ensure energy efficiency of the protocols. Channel modeling plays an important role for the optimization of communication system for WBANs. In [15], path loss is defined as:

$$PL(d) = P(d_o) + 10n\log\left[\frac{d}{d_o}\right] + \delta \quad (1)$$

where,

$$P(d_o) = 10\log\left[\frac{4\pi d_o}{\lambda}\right]^2 \quad (2)$$

Here, $d$ is the distance between transmitting and receiving node, and is variable. On the other hand, $d_o$ denotes the reference distance between a transmitting and receiving node, which is fixed. WBANs can be classified into three different scenarios, and these scenarios work well when the path loss is minimum. These three scenarios are discussed below:

### A. MAC for In-Body Communication

Developing a power-efficient MAC for In-body sensor networks is the most challenging task. In-body sensor nodes are implanted under human skin, where the signal propagation is affected by electrical properties of body that varies from person to person. Human body poses many wireless transmission challenges. Several components in a human body; thickness of tissues, their conductivity, permeability etc, differ in each human body. The main purpose of the In-body sensor nodes is to monitor In-body parameters to communicate with the other implanted devices, such as pacemakers etc [16].

Development process of a power-efficient MAC protocol for In-body sensor networks is affected by the diverse nature of In-body nodes and with electrical properties of human body. Pacemaker and Capsular endoscope are two of the examples of In-body networks and their data rate varies from few Kbps to several Mbps. In-body sensor network requires critical traffic, low latency and high reliability [16]. The use of CSMA/CA does not provide reliable solution in this scenario due to high path loss inside the human body [17]. In In-body sensor networks heating effect caused by electromagnetic wave should also be considered. Nagamine *et al.* [18] discussed the thermal influence of BAN nodes using different MAC protocols.

Fig. 1. shows path loss between two implanted sensor nodes for varying distance and frequencies. It shows that path loss increases with increasing distance between sensor nodes. In In-body communication, path loss is greater, as compared to the scenarios discussed below because human body is composed of tissues and organs, which increase path loss. Simulations are performed by taking different frequencies ranging from *800MHz* to *2800MHz*. The simulation results show that path loss between the two sensor nodes increases when frequency is increased.

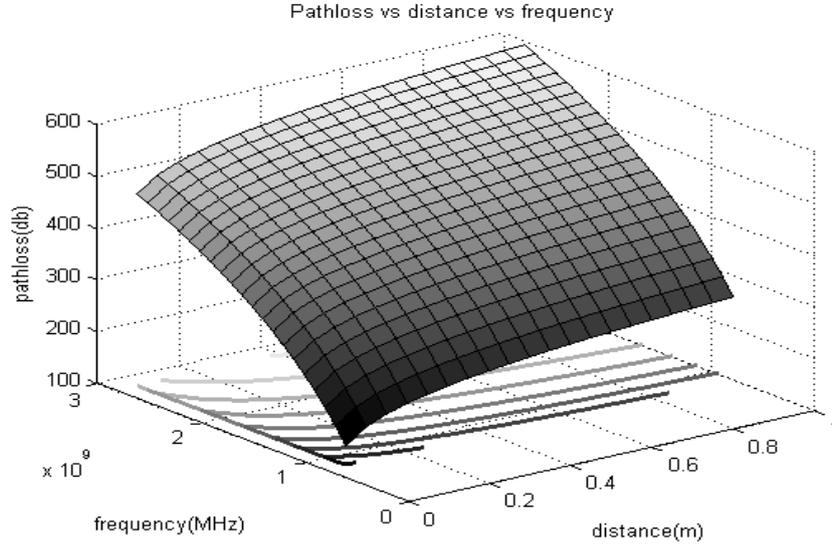

**Fig. 1.** In-Body Communication

*B. MAC for On-Body Communication*

On-body sensor networks comprise of miniaturized and non-invasive sensor nodes that are used for various applications, ranging from medical to interactive gaming and entertainment applications. Wireless Medical Telemetry Services (WMTS), unlicensed Industrial, Scientific and Medical (ISM), and Ultra-Wide Band (UWB) are some of the bands used for data transmission in On-body networks. WMTS is a licensed band designated for medical telemetry system. Due to fewer interfering sources Federal Communication Commission (FCC) urges the use of WMTS for medical applications. However, only authorized users such as physicians and trained technicians are eligible to use this band. Furthermore, restricted WMTS $14\, MHz$ band cannot support video and voice transmission. The alternative band for medical applications is $2.4\, GHz$. The band includes GBs to protect adjacent channel interference [16]. The design and implementation of a power-efficient MAC protocol for On-body sensor networks have been an emerging research topic for the last few years. H-MAC, a novel TDMA protocol for On-body sensor network exploits bio signal features to perform TDMA synchronization and improves energy efficiency [19].

Simulation results in Fig. 2 portrayed that of "On body communication", for varying distances and frequencies. Sensors are placed on human body. Simulations result depicted that increase in distance between two nodes consequently increase path loss. Path loss is less in On-body communication as compared to In-body communication because On-body sensors communicate in air and path loss is minimum in air medium. It is also observed in **Fig. 2** that by increasing frequency, path loss is also increased.

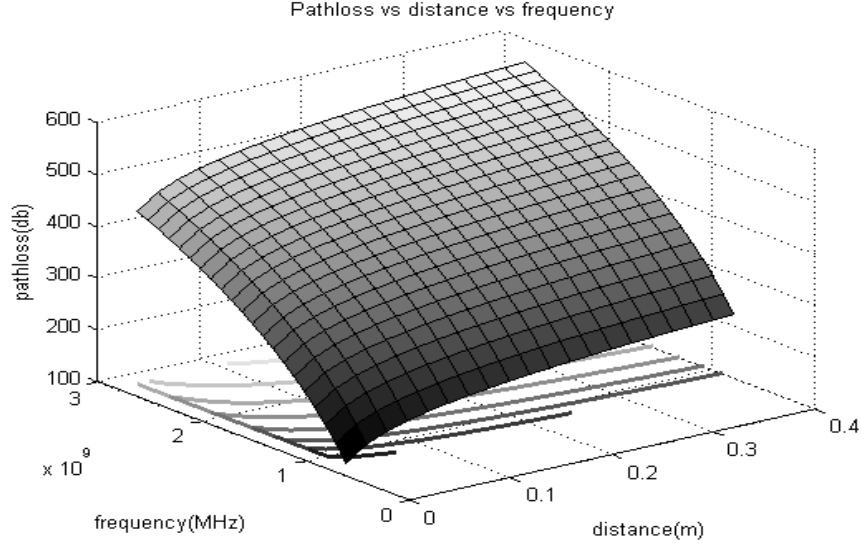

**Fig. 2.** ON-BODY Communication

*C. MAC for Off-Body Communication*

This model describes variations of the channel with respect to the following three aspects: distance between body and access points (receiver) is denoted by $\rho$, body orientation angel $\theta_A$, and transmitter based azimuth angel $\theta_{AB}$. Human body is taken as a cylinder of average size. The model uses two types of the co-ordinates i.e., Principle Coordinates (PCs) which includes cartesian coordinates, and BCS which are body cartesian coordinates.

Path loss is defined in [21] as:

$$L(\rho, \theta_A) = \begin{cases} L_o(\rho) - n_\theta(\theta_A - \theta_o) & 0 \leq \theta_A \leq \theta_{AL}(\rho) \\ L(\theta_{AL}) - n_\theta(\theta_A\{\theta_A - \theta_{AL}(\rho)\} & \theta_{AL}(\rho) \leq \theta_A \leq \theta_A S(\rho) \\ L(\theta_A S) - n_\theta(\theta_A\{\theta_A - \theta_{AL}(\rho)\} & \theta_A S(\rho) \leq \theta_A \leq \pi \end{cases} \quad (3)$$

Where, in the equation above, $\theta_{AL}$ denotes the first breakpoint angle which is observed in the lit region of the transmitter. Similarly, $\theta_{AS}$ denotes the shadow-region breakpoint angle which is observed in the shadow region of the transmitter. $n_\theta(\theta_A) = \frac{\partial L(\rho,\theta)dB}{\partial \theta}$, is the azimuth decay coefficient. $L(\rho_o)$ represents the channel loss, $L(\rho_o) = L_o(\rho_o) = 10n_p \log_{10}[\frac{\rho}{\rho_o}]$ and $L(\rho_o)$= refrence path loss. The path loss along the $\varphi$ coordinate is given as:

$$L\varphi = L(\rho, (\varphi + \varphi_A)) \quad (4)$$

Due to the symmetry of the human body with respect to z-axis
$$L(\rho, \theta_A) = L(\rho, 2\pi - \theta_A) \quad (5)$$

If the transmitter is in the LoS with the sensor at the body then path loss is low; with the movement of body the rotation angel varies and path loss increases. Fig. 3**.** shows the relationship between the distance, path loss and frequency in Off-body communication. We considered that the sensors are in LoS. Path loss is low when the sensors are LoS and high when the sensors are in NLoS. In general, path loss is minimum as compared with the scenarios discussed above.

MAC protocols are operated on either Beacon enable or CSMA/CA mode. The effect of these modes on MAC protocol with respect to power consmption is discussed in the following section.

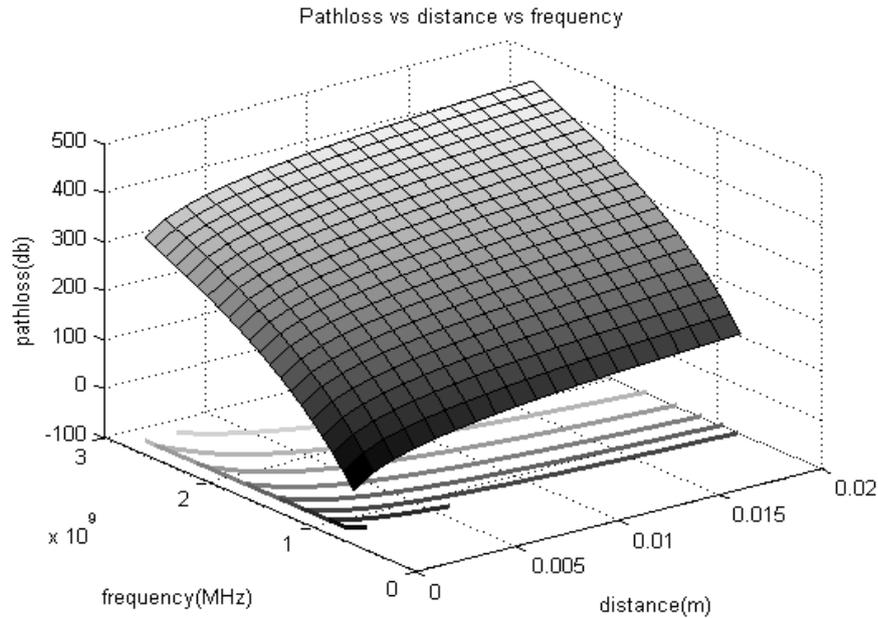

**Fig. 3.** OFF-BODY Communication

## VI. COMPARISON OF POWER MODEL FOR CSMA/CA AND BEACON MODE FOR WBASNs

In [20], Shihengcheng *et al* developed basic power consumption models i.e., CSMA/CA mode and beacon mode and compared it. They also suggested a hybrid mode to achieve the advantages of proposed schemes. Collision in WBASNs are divided into two classes i.e., InteR Network Collisions (IRNC) and IntRa Network Collision (IANC). Some general parameters are defined for the development of model are no. of WBANs (G), no. of sensors per network (N) and no. of slots in one contention period (L). In **Fig. 4**, MODEL OF CSMA/CA is potrayed.

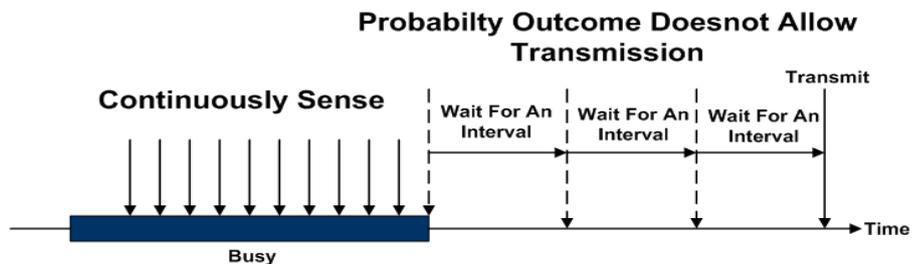

**Fig. 4.** Model of CSMA/CA

*D. CSMA/CA and beacon mode*

CSMA/CA might need more than one attempts and back-off mechanism to avoid collision. Each node repeatedly sense channel until the idle channel exists. If the channel is busy, the node back-off to the next contention period. Fig. 5 describes the flow how that a channel is sensed and uses back-off procedure. In this scenario energy spent per packet is the addition of "energy spent on packet transmission'' and "extra energy cost for channel detection''.

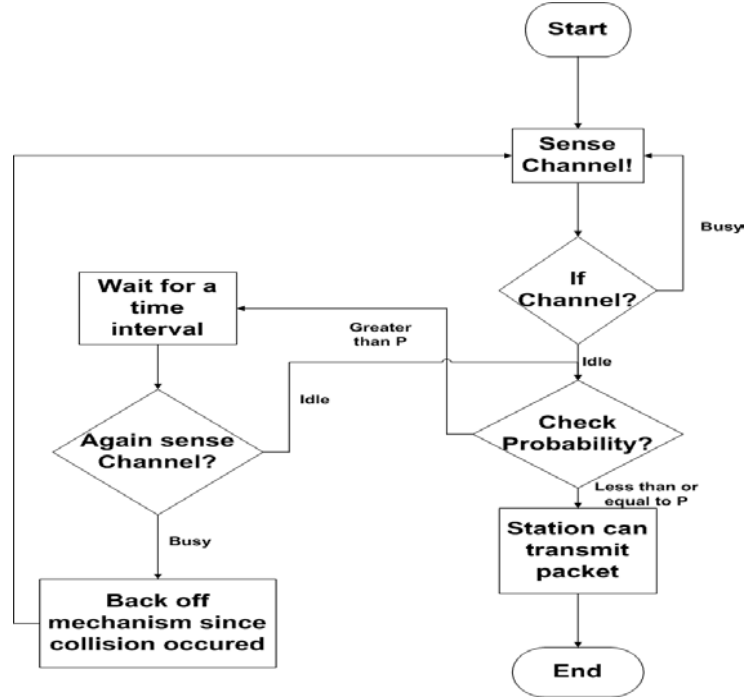

**Fig. 5.** Channel availability in CSMA/CA

$$J_{cs} = (J_{PKT} + J_{cs} - cost).Attempts_{cs} \tag{6}$$

Whereas, $Attempts_{cs}$ is the value of expected no. of contentions before a successful transmission. $P_{IANC}$ and $P_{IRNC}$ shows the probability of intra and inter network collision respectively, and are related to $Attempts_{cs}$ as following [17].

$$Attempts_{cs} = \sum_{i=1}^{\infty} i(1 - P_{IRNC} - P_{IANC})(P_{IRNC} + P_{IANC})^{i-1} \tag{7}$$

where, $i$ denotes the number of attempts. The attempt is repeated until channel is idle.

$$P_{IANC} = 1 - (1 - \tfrac{1}{L})^{N-1} \tag{8}$$

By assuming that each node uniformly competes with the rest of N-1 nodes in the same WBAN/WSN; $P_{IANC} \approx \frac{N-1}{L} \quad for\ L \gg 1$

Similarly,

$$P_{IRNC} = 1 - (1 - \tfrac{1}{L})^{(G-1)N} \tag{9}$$

$$P_{IRNC} \approx \frac{(G-1)N}{L} \qquad for\ L \gg 1 \tag{10}$$

$$Attempts_{CS} = \frac{1}{(1-P_{IRNC}-P_{IANC})} \tag{11}$$

Putting the value of $P_{IANC}$ and $P_{IRNC}$

$$Attempts_{CS} = \frac{1}{[1-(\frac{N-1}{L}+\frac{(G-1)N}{L})]} \tag{12}$$

$$Attempts_{CS} = \frac{1}{[1-\frac{GN-1}{L}]} \quad ; 1 \leq G, N \tag{13}$$

Energy spent per packet = $J_{BCN}$

$$J_{BCN} = J_{PKT} + Attempts_{BCN} \tag{14}$$

In beacon mode, the intra network collision are avoided by allocating separated time slot for different WBAN/WSN, hence in beacon mode $P_{IANC} = 0$.

$$J_{BCN} = (J_{PKT} + J_{BCN} - cost). Attempts_{BCN} \tag{15}$$

$$J_{BCN} = (J_{PKT} + J_{BCN-cost}) \sum_{i=1}^{\infty} i(1 - P_{IRNC})(P_{IRNC})^{i-1} \tag{16}$$

$$J_{BCN} = (J_{PKT} + J_{BCN-cost}) \frac{1}{\left[1-\frac{(G-1)N}{L}\right]} \quad ; 1 \leq G, N \tag{17}$$

Next, we want to relate the $J_{BCN}$ and $J_{CS}$ with the throughput of WSN. We assume that all packets have the same fixed size. For instance, in beacon mode, $Attempts_{BCN}$ implies the number of transmission attempts per packet. Thus, the throughput of *WSN* in the beacon mode can be expressed as:

$$Throughput_{BCN} = \frac{1}{L.Attempts_{BCN}} \tag{17}$$

Similarly, the throughput of CSMA/CA mode is:

$$Throughput_{CS} = \frac{1}{L.Attempts_{CS}} \tag{18}$$

Throughput is defined as the no. of packets successfully transmitted per slot time. By relating throughput to energy spent for the above two modes; the facts observed are as following:
a) CSMA/CA mode has lower power consumption than beacon mode but its capacity is limited

i.e total no. of WBAN's with the requirement of the specific throughput.
b) CSMA/CA mode adopts channel detection to avoid the collision and cost of the channel detection is lower than packet collision.

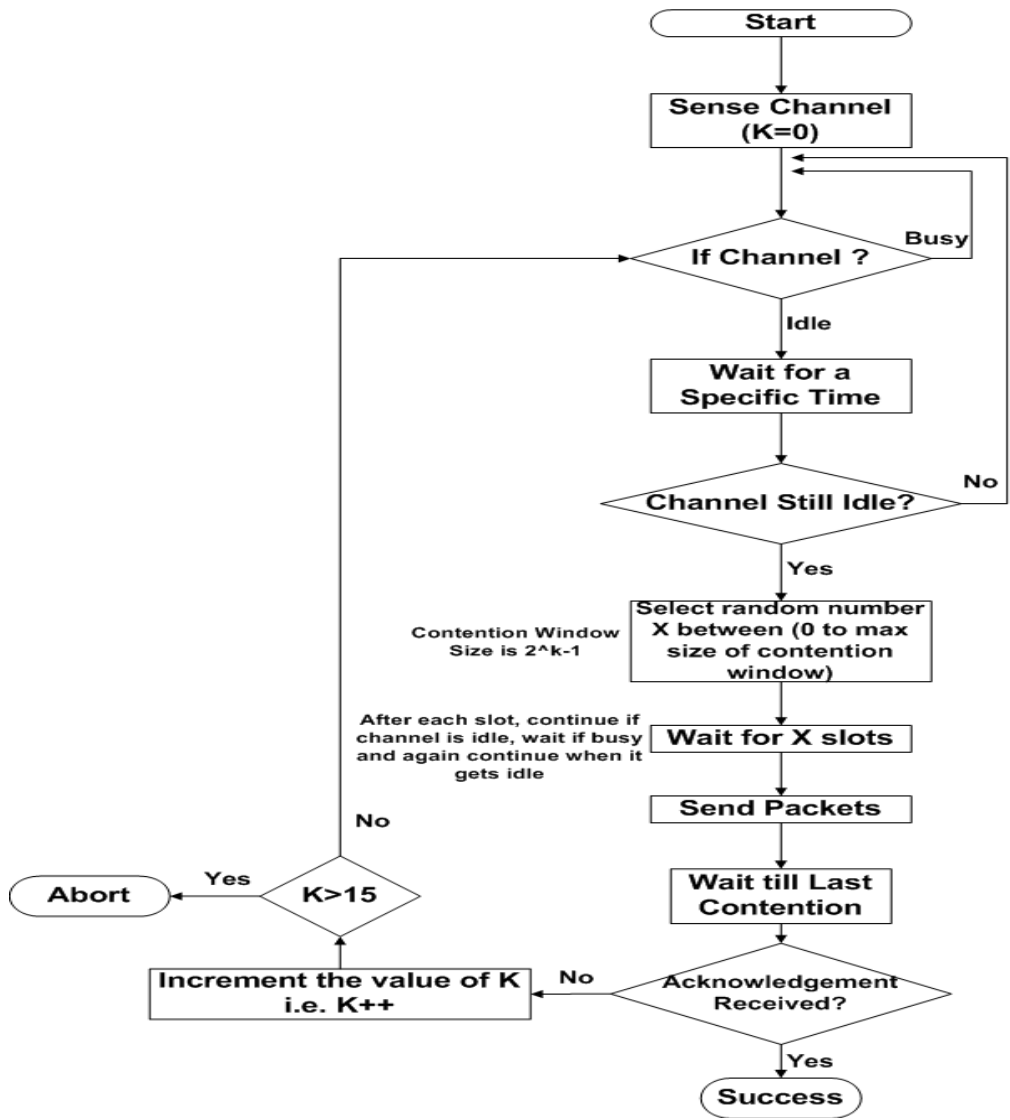

**Fig. 6.** Flow of CSMA/CA

**Figure. 6**. shows packet flow transmission through the channel. Note that the channel need to be sensed before and after Inter-Frame Space (IFS). The channel also needs to be sensed during contention time. For each slot of the contention window, the channel is sensed. If channel is detected as busy, timer stops and transmission of data starts when channel become idle.

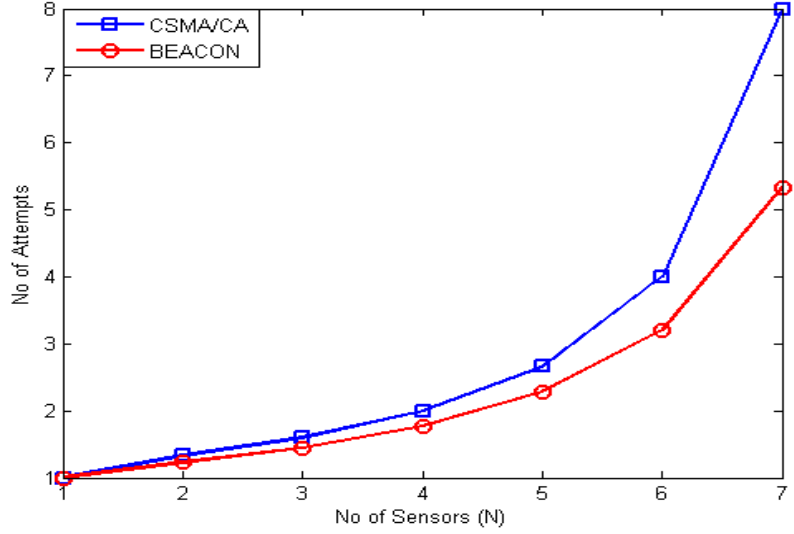

**Fig. 7.** Number of Attempts

Relationship between "no. of attempts'' and "no. of sensors'' in CSMA/CA and beacon mode is described in **Fig. 7**. The graph shows that, with the increase in the no. of sensors, no. of attempts in CSMA/CA mode increases sharply, as compared to beacon mode which has a gradual increase.

The average power consumption is computed by dividing the energy per packet by the transmission time per packet. We assume that the energy only consumed when WSN in Transmission (TX) and Reception (RX) periods in both modes.

Power consumption in beacon mode is given by:

$$\overline{W_{BCN}} = \frac{J_{BCN}}{L.Attempts_{BCN}.Time\ Per\ Slot} \tag{19}$$

$$\overline{W_{BCN}} = \frac{(J_{PKT}+J_{BCN-cost})Attempts_{BCN}}{L.Attempts_{BCN}.\frac{B_{PKT}}{R_{PHY}}} \tag{20}$$

$$\overline{W_{BCN}} = W_{ON} \frac{B_{PKT}+B_{BCN-cost}}{B_{PKT}} \frac{2\frac{Tp_{BPS}}{R_{PHY}}}{1+\sqrt{1-4\frac{Tp_{BPS}}{R_{PHY}}(G-1)N}} \tag{21}$$

Similarly, power consumption in CSMA/CA mode is given by:

$$\overline{W_{CS}} = W_{ON} \left( \frac{B_{PKT}+B_{BCN-cost}}{B_{PKT}} \frac{2\frac{Tp_{BPS}}{R_{PHY}}}{1+\sqrt{1-4\frac{Tp_{BPS}}{R_{PHY}}(GN-1)}} -(GN-1)(\frac{2\frac{Tp_{BPS}}{R_{PHY}}}{1+\sqrt{1-4\frac{Tp_{BPS}}{R_{PHY}}(GN-1)}})^2 \right) \tag{22}$$

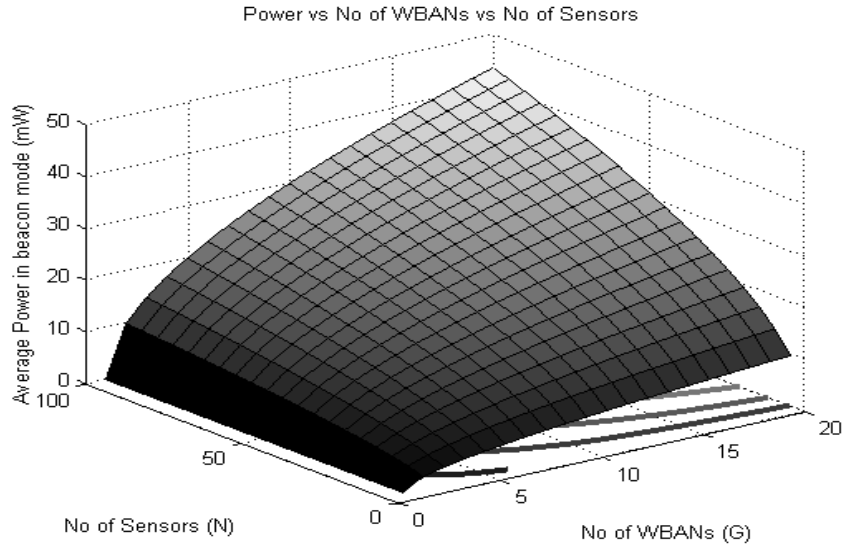

**Fig. 8.** Power consumption in Beacon mode

Simulation results of "Power in beacon mode", for varying "no of sensors" and "no of WBANs" are shown in Fig. 8. Sensors are placed on human body. Simulation results depicts that increase in "no of sensors" and "no of WBANs" increases the power. Figure. 9, shows the simulation results of "Power in CSMA/CA mode", for varying "no of sensors" and "no of WBANs". Sensors are placed on human body. Simulation results depicted that increase in "no. of sensors" and "no. of WBANs" increases the power slowly, because CSMA/CA adopted the channel detection to avoid collision. The cost of channel detection is lower than the packet collision because the detection will not listen to the channel for whole time of packet receiving. When "no. of sensors" and "no. of WBANs" are small in number power in CSMA/CA mode is almost zero.

Although CSMA/CA has the lowest power consumption, it might fail to meet high user capacity criterion. To provide both high user capacity and low power WBAN, a new access method is developed.

*A. Hybrid mode*

To achieve the goal of power consumption and the high user capacity, hybrid mode is suggested. The hybrid mode uses the beacon mode to overcome IANC and channel detection to handle IRNC. Although hybrid mode costs more power than CSMA/CA because it adopts both beacon and channel detection, hybrid mode is still a better access method than CSMA/CA when the required user capacity is high.

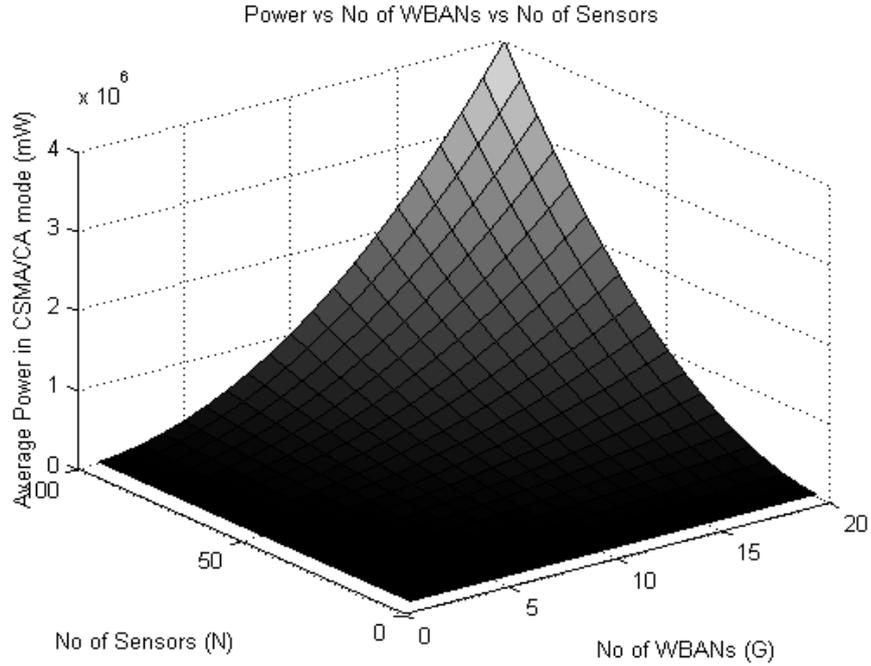

**Fig. 9.** Power consumption in CSMA/CA mode

## VII. CONCLUSION

In this study, we present an analytical survey of different MAC protocols with respect to energy efficiency and their advantages and disadvantages in WBANs. LPL scheduled contention and TDMA are the techniques used for MAC protocols. It is concluded from analytical discussion that TDMA is more power efficient however, suffers with synchronization sensitivity. Techniques for collision avoidance of different MAC protocols are comparatively analyzed in this work. Path loss model for In-body, On-body and Off-body communication in WBANs is also described. From analytical simulations, it is obsereved that path loss is maximum in In-body communication, as compare to On-body and Off-body communication because human body is composed of tissues and organs in which communication is difficult and thus results in high path loss. On-body and Off-body also show some variations in results when the source and destination sensors or nodes are LoS and NLoS. Path loss increases considerably when the sensors are NLoS. Moreover, off-body communication describes the variations of the channel with respect to the following three aspects: distance between body and access points (receiver), body orientation angel, and transmitter based azimuth angel. These results are more accurate if the human body is considered spherical instead of cylindrical. For studying power optimization techniques in WBASNs, a mathematical model for CSMA/CA and beacon mode is presented. To validate this model, we perform simulations in MATLAB. Simulation results depict that increase in "no. of sensors" and "no. of WBANs" increases the power slowly, because CSMA/CA adopts the channel detection to avoid collision. The cost of channel detection is lower than the packet collision because the detection will not listen to the channel for whole time of packet receiving and "Power in beacon mode", for varying "no of sensors" and "no. of WBANs". Moreover, that increase in "no. of sensors" and "no. of WBANs" increases the power is also analysed from simulations.

# VIII. REFERENCES

1. Samira Afzal, 2012. Localization Algorithm for Large Scale Mobile Wireless Sensor Networks: J. Basic. Appl. Sci. Res., 2(8): 7589-7596.
2. Amjad Osmani, 2012. Design and Evaluation of New Intelligent Sensor Placement Algorithm to Improve Coverage Problem in Wireless Sensor Networks: J. Basic. Appl. Sci. Res., 2(2): 1431-1440.
3. Vahid Vahidi et al., 2012. Evaluation of Several Anchor Placement Scenarios and Positioning Methodologies in Wireless Sensor Networks: J. Basic. Appl. Sci. Res., 2(11): 10945-10950.
4. Anand Gopalan, S. and Park, J.T., 2010. Energy-efficient MAC protocols for wireless body area networks: Survey: ICUMT.
5. Ali Barati, Ali Movaghar, Samira Modiri, Masoud Sabaei, 2012. A Reliable & Energy-Efficient Scheme for Real Time Wireless Sensor Networks Applications: J. Basic. Appl. Sci. Res., 2(10): 10150-10157.
6. Kutty, S. and Laxminarayan, JA., 2010. Towards energy efficient protocols for wireless body area networks, ICII,.
7. Ullah, S. and Shen, B. and Riazul Islam, SM and Khan, P. and Saleem, S. and Sup Kwak, K., 2009. A study of MAC protocols for WBANs: SENSOR.
8. Timmons, NF and Scanlon, WG., 2009. An adaptive energy efficient MAC protocol for the medical body area network: VITAE.
9. Omeni, O. and Wong, A. and Burdett, A.J. and Toumazou, C, 2008. Energy efficient medium access protocol for wireless medical body area sensornetworks: IEEE.
10. Marinkovic, S.J. and Popovici, E.M. and Spagnol, C. and Faul, S. and Marnane, W.P., 2009. Energy-efficient low duty cycle MAC protocol for wireless body area networks: IEEE.
11. Fang, G. and Dutkiewicz, E., 2009. BodyMAC: Energy efficient TDMA-based MAC protocol for wireless body area networks: ISCIT,.
12. Ullah, S. and Kwak, K., 2010. An ultra low-power and traffic-adaptive medium access control protocol for wireless body area network: J. Med. Syst,.
13. W. Ye, J. Heidemann, and D. Estrin, 2002. An energy-efficient MAC protocol for wireless sensor networks. In Proceedings of the IEEE Infocom, pp: 1567-1576.
14. T. Van Dam and K. Langendoen, 2003. An adaptive energy-efficient MAC protocol for wireless sensor networks. In ACM Conference on Em-bedded Networked Sensor Systems (Sensys), pp: 171-180.
15. http://en.wikipedia.org/wiki/Path loss.
16. Sana Ullah, Xizhi An, and Kyung Sup Kwak. 2009. Towards Power Effi-cient MAC Protocol for In Body and On-Body Sensor Networks, A. Hkansson et al. (Eds.): KES-AMSTA, LNAI 5559, pp: 335345.
17. Shu, F., Dolmans, G., 2008: QoS Support in Wireless BANs. IEEE P802.15 Working Group for Wireless Personal Area Networks, WPANs.
18. Zhen, B., Li, H.-B., Kohno, R., 2008. IEEE body area networks and medical implant communications. In: Proceedings of the ICST 3rd International Conference on Body Area Networks, Tempe, Arizona.
19. Huaming Li, H., Jindong Tan, H.,2007. Heartbeat Driven Medium Access Control for Body Sensor Networks, In: Proceedings of the 1st ACM SIGMOBILE international workshop on Systems and networking sup-port for healthcare and assisted living environments, pp: 2530.
20. Goulianos, A.A. and Brown, T.W.C. and Stavrou, S., 2008. Novel Path-Loss Model for UWB Off-Body Propagation: VTC Spring.
21. Ullah, S. and Kwak, K.S., 2010. An ultra low-power and traffic-adaptive medium access control protocol for wireless body area network: Journal of Medical Systems.